\documentclass[preprint,preprintnumbers,amsmath,amssymb,superscriptaddress]{revtex4}
\usepackage{txfonts}% Physical Review B
\usepackage{graphicx}% Include figure files
\usepackage{dcolumn}% Align table columns on decimal point
\usepackage{bm}% bold math
\usepackage{array}
\usepackage[colorlinks=true,plainpages=false,linkcolor=blue,urlcolor=blue,citecolor=blue,pdfpagemode=UseNone,pdfstartview=FitBH]{hyperref}

\begin{document}

\title{ Nodal superconducting gap in LiFeP revealed by NMR: contrast with LiFeAs}
\author{A. F. Fang}
\affiliation{Derpartment of Physics, Beijing Normal University, Beijing 100875, China}

\author{R. Zhou}
\thanks{Electronic address: rzhou@iphy.ac.cn}
\affiliation{Institute of Physics, Chinese Academy of Sciences,\\
 and Beijing National Laboratory for Condensed Matter Physics,Beijing 100190, China}
\affiliation{Songshan Lake Materials Laboratory, Dongguan, Guangdong 523808, China}

\author{H. Tukada}
\affiliation{Department of Physics, Okayama University, Okayama 700-8530, Japan}

\author{J. Yang}
\affiliation{Institute of Physics, Chinese Academy of Sciences,\\
 and Beijing National Laboratory for Condensed Matter Physics,Beijing 100190, China}

\author{Z. Deng}
\affiliation{Institute of Physics, Chinese Academy of Sciences,\\
 and Beijing National Laboratory for Condensed Matter Physics,Beijing 100190, China}

\author{X. C. Wang}
\affiliation{Institute of Physics, Chinese Academy of Sciences,\\
 and Beijing National Laboratory for Condensed Matter Physics,Beijing 100190, China}

\author{C. Q. Jin}
\affiliation{Institute of Physics, Chinese Academy of Sciences,\\
 and Beijing National Laboratory for Condensed Matter Physics,Beijing 100190, China}

\author{Guo-qing Zheng}
\affiliation{Institute of Physics, Chinese Academy of Sciences,\\
 and Beijing National Laboratory for Condensed Matter Physics,Beijing 100190, China}
\affiliation{Department of Physics, Okayama University, Okayama 700-8530, Japan}

\date{\today}% It is always \today, today,
             %  but any date may be explicitly specified

\begin{abstract}
{Identifying the uniqueness of FeP-based superconductors may shed new lights on the mechanism of superconductivity in iron-pnictides. Here, we report nuclear magnetic resonance(NMR) studies on LiFeP and LiFeAs which have the same crystal structure but different pnictogen atoms. The NMR spectrum is sensitive to inhomogeneous magnetic fields in the vortex state and can provide the information on  the superconducting pairing symmetry through the  temperature dependence of London penetration depth
$\lambda_L$. We find that $\lambda_L$ saturates below $T \sim 0.2$ $T_c$  in LiFeAs, where $T_c$ is the superconducting transition temperature, indicating nodeless superconducting gaps. Furthermore, by using a two-gaps model, we simulate the temperature dependence of $\lambda_L$ and obtain the superconducting gaps of LiFeAs, as $\Delta_1 = 1.2$ $k_B T_c$ and $\Delta_2 = 2.8$ $k_B T_c$, in agreement with previous result from spin-lattice relaxation. For LiFeP, in contrast, the London penetration depth $\lambda_L$ does not show any saturation down to $T \sim 0.03 $ $T_c$, indicating nodes in the superconducting energy gap function.  Finally, we demonstrate that the strong spin fluctuations with diffusive characteristics exist in LiFeP, as in some cuprate high temperature superconductors.
}
\end{abstract}



\maketitle

%The presence of the pseudogap  is still puzzling, and the understanding of its nature is important for exploring the mechanism of high-temperature superconductivity in cuprates.

\section{Introduction}\label{section1}

As the second class of high-temperature superconductors, iron-based superconductors were discovered more than one decade ago\cite{Kamihara2008}. But its superconducting pairing mechanism is still unclear. The phase diagram of iron-pnictides is very similar to the cuprates high-temperature superconductor family\cite{Keimer2015}.
Antiferromagnetism and nematic orders exist around the superconducting dome and both compete with superconductivity in these two families\cite{Dai2015}.
%Antiferromagnetism and nematic orders that compete with superconductivity in these two families, are observed around the superconducting dome\cite{Dai2015}.
Therefore both spin and nematic fluctuations are suggested to be candidates for the glue of the superconducting pairing\cite{Fernandes2014}. Although, the symmetry of the superconducting gap in cuprates is believed to be $d$-wave \cite{Shen1993,Wollman1993}, the situation is more complicated for the iron-pnictides.
 %Firstly, two gaps were observed below the transition temperature, instead of a single superconducting energy gap.
Firstly, there are multiple superconducting gaps as first evidenced by the spin-lattice relaxation rate and the Knight shift\cite{Matano2008}, instead of a single superconducting energy gap in the cuprates.
Secondly, the gap symmetry is different in different iron-based families. In FeAs-based superconductors, the superconducting gap is found to be isotropic and fully-opened\cite{Ding2008,Li2011,Oka2012,Ge2013}. But for P-doped BaFe$_2$As$_2$ which is equivalently doped, previous studies suggest the existence of nodes in the superconducting gaps\cite{Zhang2012,Yoshida2014}. If As is completely substituted by P, the  pnictogen height above the iron plane will become smaller, which is suggested to be an important factor in the theory based on spin fluctuations\cite{Kuroki2009}. Therefore, clarifying the uniqueness of FeP-based superconductor can shed lights on the mechanism of superconduting pairing in iron-pnictides.

LiFeP is a superconducting material with transition temperature $T_c\sim$ 4.2 K\cite{Deng2009}. Its crystal structure is identical to LiFeAs($T_c\sim$ 18 K)\cite{Wang2009}, but the height of the P site is much smaller than that of As. Previous tunnel diode oscillator (TDO) measurements found that the London penetration depth shows a flat temperature dependence in LiFeAs but a linear temperature dependence in LiFeP, suggesting nodeless and nodal superconducting gap respectively\cite{Hashimoto2012}. However, TDO measurement is only sensitive to the change of penetration depth on the surface of the sample which can be affected by disorder or lattice distortion from the surface. Until now, no bulk measurement on the London penetration depth $\lambda_L$ has been done. NMR spectrum is sensitive to inhomogeneous magnetic fields in the vortex state, from which $\lambda_L$ of the bulk sample can be directly deduced\cite{Wang2018,Oh2013}.
%It is also reported that spin fluctuations are strongly enhanced towards $T_c$ in LiFeAs and some Fe-based families.
Besides the properties of the superconducting state, the electron correlations in the normal state of LiFeP are also of interest. In most iron-based superconductors, strong spin fluctuations have been observed\cite{Ning2010,Oka2012,Nakai2010,Zhou2013}, and quantum critical point related to the magnetic order is suggested inside the superconducting dome\cite{Hashimoto2012-2,Wang2018}. Unlike these compounds, a previous NMR study at $B_0$ = 4.65 T has suggested that low-energy spin fluctuations are very weak in LiFeP\cite{Man2014}. However, for spin-lattice relaxation rate 1/$T_1$ measurement, the NMR frequency, which is related to the energy of spin fluctuations, can play an important role. Therefore, 1/$T_1$ measurements at different fields are needed in order to investigate the intrinsic spin fluctuations.
%So it is important to investigate the superconducting gap symmetry and spin dynamics by NMR at very low fields.
%In addition, LiFeAs, the counterpart of LiFeP, attracts a lot of attention recently.
%Tunable vortex Majorana zero modes are observed in LiFeAs superconductor\cite{Ding2020}, which offers a promising platform for topological superconductivity and quantum computation.

In this work, we investigate the superconducting gap symmetry of LiFeP and LiFeAs by detailed NMR studies of London penetration depth $\lambda_L$. Nodal superconductivity is revealed in LiFeP while LiFeAs is found to be a nodeless superconductor. For LiFeP, strong spin fluctuations with diffusive characteristics are found by spin-lattice relaxation measurements, which is similar to some cuprate superconductors.

\section{Experiment}

The LiFeAs single crystal and LiFeP polycrystal samples were grown by the solid-state reaction and the self-flux methods, respectively\cite{Deng2009,Wang2009}. The $^{75}$As spectra were obtained by integrating the spin echo as a function of frequency at 7.5 T. The $^{31}$P-NMR spectra were obtained by fast Fourier transform of the spin echo. The pulse width is only 5 $\mu$s in order to cover the full spectrum.  The $T_1$ was measured by using the saturation-recovery method, and obtained by a good fitting  of the nuclear magnetization to  $1-M\left(t\right)=M_0 e^{-t/T_1}$, where $M(t)$ is the nuclear magnetization at time $t$ after the single saturation pulse and $M_0$ is the nuclear magnetization at thermal equilibrium.

\section{Results and Discussion}

\subsection{London penetration depth in LiFeP: contrast with LiFeAs}

%In order to show the uniqueness of LiFeP, we also studied the temperature-dependent $\lambda _{L}^{-2}$ in LiFeAs.
Figure 1(a) shows $^{75}$As-NMR central line at various temperatures, which can be well fitted by a single Lorentz function. By cooling down into the superconducting state, the NMR line shifts to lower frequency and broadens almost symmetrically. In the vortex state, the magnetic field $B_0$ penetrates into a sample in the unit of quantized flux $\phi_0$ = 2.07 $\times$ 10$^{-15}$ T m$^2$, thus the field becomes inhomogeneous, leading to the observed broadening in Fig. 1(a). The shift of spectrum is due to the singlet pairing and diamagnetism from the vortex-lattice formation. For $B_{c1} \ll B_0 \ll B_{c2}$, where $B_{c1}$ and $B_{c2}$ are the lower and upper critical field respectively, the field distribution $\Delta B$ can be written as\cite{Brandt}:

\begin{equation}
\Delta B=0.0609\frac{{{\phi }_{0}}}{\lambda _{L}^{2}},
\label{dB}
\end{equation}

\begin{figure}
\includegraphics[width=12cm]{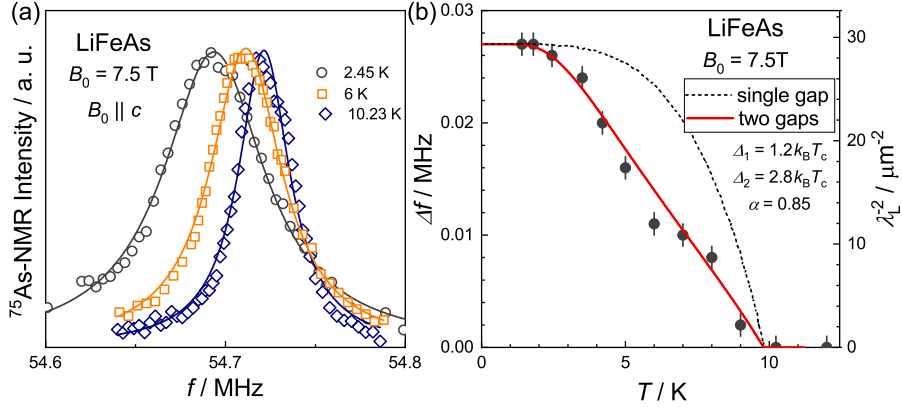}
\caption{(color online) (a) $^{75}$As-NMR spectra of LiFeAs at various temperatures with magnetic field B$_0$ = 7.5 T applied along $c$-axis. The spectra are fitted by a single Lorentz function. (b) Temperature dependence of the line broadening $\Delta f$ and London penetration depth $\lambda _{L}^{-2}$ of LiFeAs. The black dashed curve represents the variation with temperature expected for a conventional s-wave superconductor\cite{Sonier}.
The red solid curve represents the simulation by the two-gaps model described in the text.
}
\label{spec}
\end{figure}

which can be detected by the NMR spectrum broadening $\Delta f = \gamma_n \Delta B$ , where $\gamma_n$ is the gyromagnetic ratio. In both normal state and superconducting state, the spectra can be well fitted by a single Lorentz function. Theoretically, the NMR lineshape in the superconducting state should be asymmetric due to inhomogeneous distribution of the magnetic field. However, the shape we observed is rather symmetric, which is in agreement with the previous NMR study in NaFe$_{1-x}$Co$_x$As\cite{Oh2013,Wang2018}.
%The first possible explanation is a random pinning of the vortex lattice by defects or domain boundaries. This will contribute an additional Gaussian distribution, and then the lineshape will be symmetric.
%This could be the case for LiFeP, since polycrystalline sample is used for measurements.
This might be because the vortex-cores have small random displacements from triangular lattice in a 2D layered system when the correlation between different layers is small. Such displacements will broaden the effective core radius and truncate the high-field tail in the field distribution. Then the line will become more symmetric, like the case in Bi$_2$Sr$_2$CaCu$_2$O$_{8+\delta}$\cite{Harshman1993}. This can explain why the broadening in LiFeAs is rather symmetric, since iron-based superconductors are also quite 2D.
%Although this effect will modify a little bit of the pre-factor in Eq. 1, the main conclusion in our study will not be affected.

The full width at half maximum (FWHM) of a convolution of two Lorentzian functions is the sum of individual FWHMs, so the broadening can be obtained by simply subtracting the $T$-independent width above $T_c$, $\Delta$$f$ = FWHM($T$) - FWHM($T$~$>$~ $T_{c}$). In Fig. 1(b), we summarize the temperature dependence of  $\Delta$$f$ and $\lambda _{L}^{-2}$ which start to saturate below $T \sim$ 0.2 $ T_c$. By using Eq. (1), $\lambda _{L}(T \rightarrow 0)$ = 185 nm is calculated , which is consistent with the result, $\lambda _{L}$ = 210 nm, obtained by  small-angle neutron scattering (SANS)\cite{Inosov2010}. In the London theory, $\lambda _{L}^{-2}$ is proportional to  the superconducting carrier density $n_s$ as\cite{Carrington}:

\begin{equation}
n_s=\frac{m^\ast c^2}{4\pi e^2{\lambda^2}_L},
\label{ns}
\end{equation}

where $m^{\ast}$ is the effective mass of the carriers. When the superconducting correlation length is much smaller than  $\lambda _{L}$, the superconducting carrier density $n_s$ can be expressed as\cite{Carrington}:

\begin{equation}
\widetilde n=1+2\int_0^\infty dE\frac{\partial f\left(E\right)}{\partial E}\frac{N\left(E\right)}{N_0}
\label{ns2}
\end{equation}

where $\widetilde n = n_s(T) / n_s (0)$. $f(E)$ is  the Fermi function. $N(E)$ and $N_0$ are the angle averaged superconducting density of states and the normal density of states value at the Fermi level respectively. For a $s$-wave superconductor, when temperature is smaller than 0.5 $T_c$, the change in the penetration depth $\lambda _{L}$ with respect to their zero temperature value is\cite{Carrington}:

\begin{equation}
\frac{\lambda\left(T\right)}{\lambda\left(0\right)}=1+\sqrt{\frac{\pi\Delta_0}{2k_BT}}\text{exp}\left(\text{-}\frac{\Delta_0}{k_BT}\right)
\label{swave}
\end{equation}

where $\Delta$ is the zero temperature value of the superconducting energy gap, and $k_B$ is the Boltzmann constant. One can immediately see that $\lambda _{L}^{-2}$ should be nearly temperature independent at low temperatures($T < 0.4 T_c$) for a conventional $s$-wave\cite{Sonier}%For a superconductor with a single $s$-wave energy gap, $\lambda _{L}^{-2}$ should saturate at $T \sim 0.4$ $ T_c$
, as shown by the dashed line in Fig. 1(b) which is distinct from our results.  We therefore simulate our results by assuming two $s$-wave gaps, $\Delta_1$ and $\Delta_2$.  If the contribution to the superfluid density for $\Delta_1$ is $\alpha$, then it will be 1 - $\alpha$ for $\Delta_2$. The total superfluid density $n_{tot}$ is $\alpha$$n_{s1}$ + (1 - $\alpha$)$n_{s2}$. From Eq. 3, the superfluid density can be further expressed as\cite{Carrington}:

\begin{equation}
\widetilde n\left(T\right)=1+2\int_\Delta^\infty\operatorname dE\frac{\partial f\left(E\right)}{\partial E}\frac E{\sqrt{E^2-\Delta^2\left(T\right)}}
\end{equation}

By this way, we simulate the temperature dependence of $\Delta f$ as shown in Fig. 1(b). The parameters  $\Delta_1$ = 1.2 $k_B$$T_c$, $\Delta_2$ = 2.8 $k_B$$T_c$, and $\alpha$ = 0.85 are obtained. The two-gaps feature in the superconducting state was also demonstrated by previous spin-lattice relaxation measurements\cite{Li2010}, in which a 'knee' behavior was observed in temperature-dependent 1/$T_1$. From the fitting by two s-wave gaps, $\Delta_1$ = 1.3 $k_B$$T_c$ and $\Delta_2$ = 3.0 $k_B$$T_c$\cite{Li2010} was obtained, which is in good agreement with the present results.
We also note that $\Delta_1$ = 1.6 $k_B$$T_c$ of hole-like Fermi surfaces and $\Delta_2$ = 2.3 $k_B$$T_c$ of electron-like Fermi surfaces were observed by a previous ARPES study in LiFeAs\cite{Borisenko2010}, which is also consistent with our study. Furthermore, the hole-like Fermi surfaces are found to be larger than the electron-like Fermi surfaces\cite{Borisenko2010}, which is in agreement with our simulation that $\alpha$ is larger than 0.5. This means that the main contribution to the quasi-particles in the superconducting state is from the smaller superconducting energy gap $\Delta_1$ that is of hole-like Fermi surfaces. Namely the superconducting energy gap on hole-like Fermi surfaces is smaller than the gap on electron-like surfaces. This is in contrast to the situation in the BaFe$_2$As$_2$ family where the superconduting energy gap on hole-like Fermi surfaces is larger\cite{Li2011}. It implies that the pairing mechanism in LiFeAs is indeed unique\cite{Li2020}. More theoretical studies in this regard is needed in the future.
%, leading to smaller hole-type ($\Delta_1$) and larger electronic-type ($\Delta_2$) energy gaps in LiFeAs.
%Since the contribution to the superfluid density for $\Delta_1$ is bigger than that of $\Delta_2$, combing with the ARPES results that hole-like Fermi surfaces are larger than electron-like Fermi surfaces\cite{Borisenko2010} , $\Delta_1$ should be of hole-like Fermi surfaces while $\Delta_2$ is of electron-like Fermi surfaces.
%Then we compare the energy gaps between LiFeAs and BaFe$_2$As$_2$ family. It turns out that the energy gap of hole-like Fermi surfaces in LiFeAs is smaller than that of electron-like Fermi surfaces,

\begin{figure}
\includegraphics[width= 12cm]{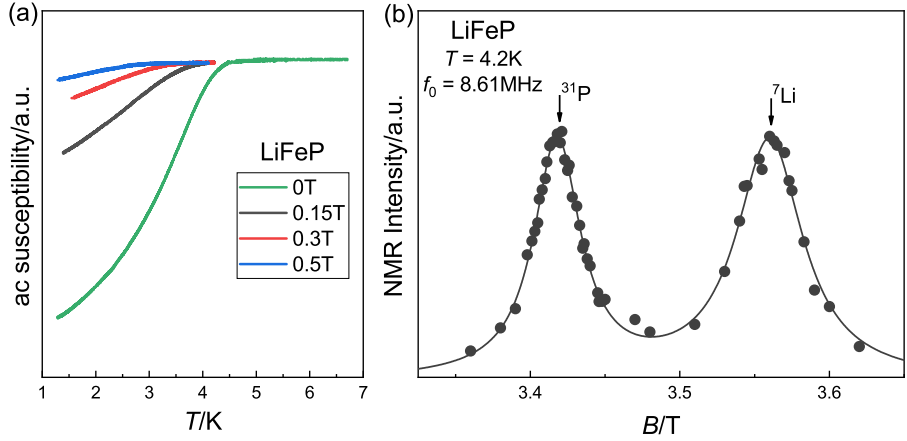}
\caption{(color online)  (a) Temperature dependence of AC susceptibility of LiFeP at various fields. (b) NMR spectrum of LiFeP obtained by sweeping the magnetic fields at 4.2 K. The solid curve is fitted by two Lorentz functions.
}
\label{Tc}
\end{figure}

\begin{figure}
\includegraphics[width= 6 cm]{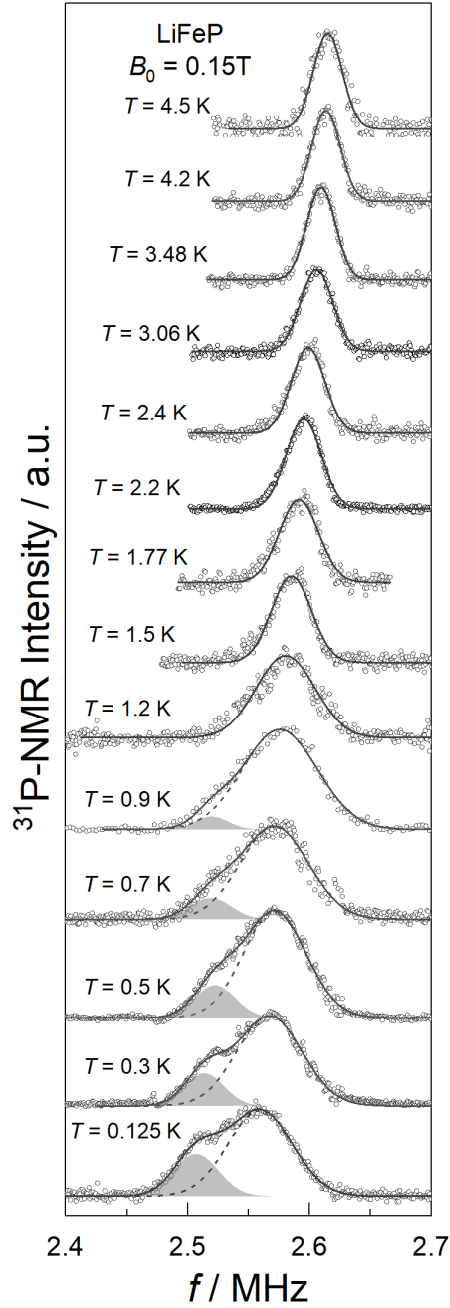}
\caption{(color online) $^{31}$P-NMR spectra of LiFeP at various temperatures with $B_0$ = 0.15 T. The spectra above $T$ = 1.2 K are fitted by a single Gaussian function, while the spectra below $T$ = 1.2 K is fitted by two Gaussian functions. The left peak (shaded area) is from $^7$Li nuclei (see text for detail).
}
\label{spec_all}
\end{figure}

Figure 2(a) shows the temperature dependence of the resonance frequency of the NMR coil at various magnetic fields. The superconducting transition temperature $T_c$ of the sample is found to be around 4.2 K at zero field, which is similar to an earlier report determined by DC susceptibility measurements\cite{Deng2009}. Fig. 2(b) shows the NMR spectrum measured at $T$ = 4.2 K by sweeping the magnetic field. We note that only one peak is observed for both $^{31}$P and $^{7}$Li nuclei. The total Hamiltonian for the nuclei with spin $I$ can be expressed as\cite{Abragam}:

\begin{equation}
\mathcal{H} = \gamma \hbar {I_z}{H_0}(1 + K) + \frac{{{e^2}qQ}}{{4I(2I - 1)}}[3(I_z^2 - {I^2})(3{\cos ^2}\theta  - 1)],
\label{dB}
\end{equation}

where $K$ is the Knight shift, $eq$ is the electric field gradient (EFG) along the principle axis $z$, Q is the nuclear quadrupole moment, and $\theta$ is the angle between the magnetic field and the principle axis of the EFG. For $^{31}$P with $I$ = 1/2, only one peak is expected. For $^{7}$Li with $I$ = 3/2, the NMR spectra should contain three lines. The fact that only one peak can be observed in our measurement is probably because the nuclear quadrupole moment $Q$ of $^{7}$Li is very small\cite{Jeglic2009,Ma2010} and the central and satellite lines overlap.

\begin{figure}
\includegraphics[width= 12cm]{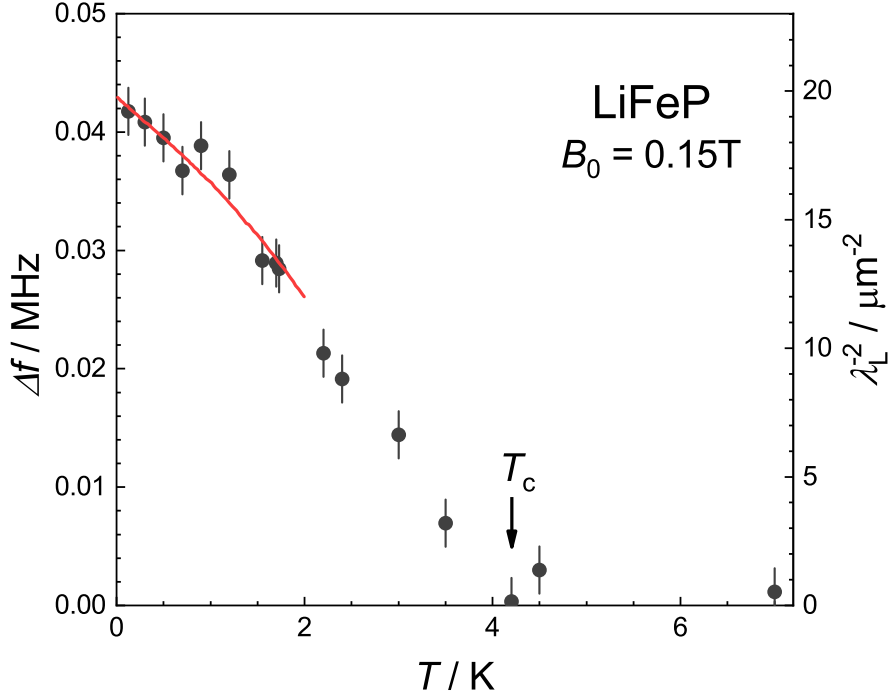}
\caption{(color online) Temperature-dependent line broadening $\Delta f$ and the London penetration depth $\lambda _{L}^{-2}$ of LiFeP. The red solid curve is the theoretical calculation based on a $d$-wave model\cite{Hirschfeld}.
}
\label{spec_P}
\end{figure}

Figure 3 shows the temperature dependence of $^{31}$P-NMR spectra at $B_0$ = 0.15 T. Similar to the observation in LiFeAs, a symmetric line broadening and a shift to the lower frequency below $T_c$ are also seen in LiFeP. Above $T$ = 1.2 K, we used a short pulse-repetition time in order to average the $^{31}$P signal as much times as possible. As a result, $^{7}$Li NMR signal is not seen, which has a long relaxation time. Below $T$ = 1.2 K, the spectrum was obtained by using a long pulse-repetition time in order to avoid heating the sample up. So a new peak is observed which is contributed from 7Li nuclei as shown in Fig. 3.
%In principle, a peak contributed from $^{7}$Li should be observed as shown in Fig. 2. However, Li is far from FeP layer, making its spin-lattice relaxation time much longer than that of $^{31}$P nuclei. To obtain enough signal to noise ratio at 0.15 T, a relatively short interval time between two measurements is used for the $^{31}$P-NMR experiment. So a peak from $^{7}$Li is not observed above $T$ = 1.2 K. Below $T$ = 1.2 K, in order not to heat the sample, a much longer interval time between two measurements was used. Then this makes the peak contributed from $^{7}$Li nuclei appear.
Since the FWHM of a convolution of two Guassian functions is the quadratic sum of individual FWHMs, the broadening can be obtained as $\Delta_f = \sqrt{\left(\text{FWHM}\left(T\right)\right)^2\text{-}\left(\text{FWHM}\left(T=T_c\right)\right)^2}$.

Figure 4 shows the $T$ dependence of the line broadening $\Delta f$. We find that $\lambda _{L}^{-2}$, being proportional to $\Delta f$, increases continuously below $T_c$ without any saturation down to the lowest temperature we have reached, $T$ = 0.125 K $\sim$  0.03 $T_c$. Such temperature dependent behavior is in contrast to the observation of LiFeAs, which can be explained by the difference of the symmetry of the superconducting energy gap. If line nodes exist in the superconducting energy gap $\Delta(\mathbf{k})$, the density of states averaged over $\mathbf{k}$ will increase linearly with energy. Therefore $\lambda _{L}^{-2}$ should be proportional to $T$ at low temperatures, which was observed by previous $\mu$SR measurements on clean polycrystalline samples of high-$T_c$ cuprate superconductors\cite{Sonier}. Fig. 4 shows that the temperature dependence of $\lambda _{L}^{-2}$ is in good agreement with the theoretical calculation based on a $d$-wave model, as shown by the red line\cite{Hirschfeld}. Our results therefore indicate that there are nodes in the superconducting energy gap. We note though that our results do not mean that the superconducting gap of LiFeP has $d$-wave symmetry since the temperature dependence of $\lambda _{L}^{-2}$ alone cannot discern $d$-wave symmetry from other nodal superconducting energy gap symmetries, such as circular line node observed in BaFe$_2$(As$_{0.7}$P$_{0.3}$)$_2$\cite{Zhang2012}.

\subsection{Spin fluctuations in LiFeP}

In most iron-pnictides, spin fluctuations have been observed in the normal state and considered as a possible glue for cooper pairs\cite{Oka2012,Nakai2010,Ning2010,Zhou2013}. However, in both LiFeP and LiFeAs, previous spin-lattice relaxation rate 1/$T_1$ measurements show that spin correlations are rather weak\cite{Li2010,Man2014}. For LiFeAs, 1/$T_1$ was measured at both zero and high fields\cite{Li2010,DaiYM2015}, indicating that the spin correlations are indeed very weak at low energies. This is consistent with the ARPES study which shows that the electron and hole pockets are mismatched, leading to the bad nesting of the Fermi surfaces and then weak spin fluctutions\cite{DaiYM2015}. However, for LiFeP, 1/$T_1$ was measured only at 4.65 T\cite{Man2014}. In order to obtain the complete information about spin dynamics, we measure 1/$T_1$ at various fields as shown in Fig. 5. At 7 T, the spin-lattice relaxation rate divided by temperature, 1/$T_1$$T$, is indeed nearly temperature independent. With decreasing field, 1/$T_1$$T$ starts to increase below $T \sim$ 10 K. At 0.15 T, a strong enhancement of 1/$T_1$$T$ is clearly observed even in the superconducting state, indicating that spin correlations become much stronger at very low energies. In La$_{2-x}$Sr$_x$CuO$_4$, 1/$T_1$$T$ also shows an enhancement with cooling in the superconducting state, which is related to the spin glass transition\cite{Frachet2020}. In such case, spin correlations should be further enhanced at higher magnetic fields due to the suppression of superconductivity. It means that 1/$T_1$$T$ should have a stronger temperature dependence at higher fields, in contrast to the observation in LiFeP.

\begin{figure}
\includegraphics[width= 12cm]{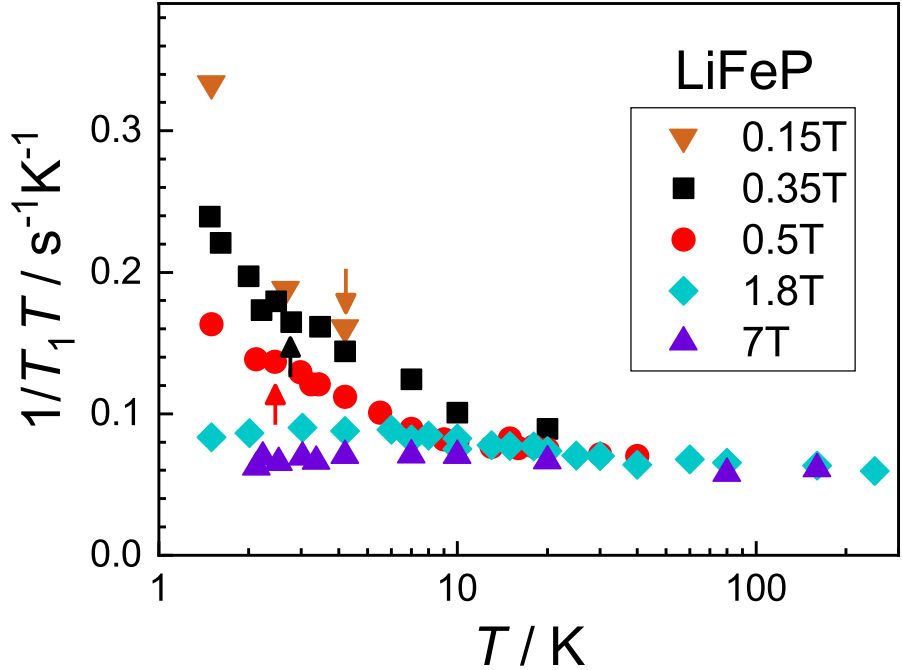}
\caption{(color online) Temperature evolution of 1/$T_1$$T$ of LiFeP at various fields. The arrows mark the onsets of superconducting transition $T_c$ under respective fields. The error bar for 1/$T_1$$T$ is the s.d. in fitting the nuclear magnetization recovery curve and is smaller than the symbol size.
}
\label{T1T}
\end{figure}

In Fig. 6(a), we plot the value of 1/$T_1$$T$ measured at 1.5 K and 4.2 K as a function of $f_0^{-1/2}$. 1/$T_1$$T$ appears to be proportional to $f_0^{-1/2}$, which is a typical behavior of the electronic spin diffusion in one dimensional(1D) systems\cite{Benner}. The possibility of two-dimensional (2D) spin diffusion where 1/$T_1$$T$ $\propto$ -ln($f$) can not be fully excluded as shown in Fig. 6(b), although the fitting for 2D is not as good as the 1D situation. In a cuprate compound Tl$_2$Ba$_2$CuO$_y$, 1/$T_1$$T$ $\propto$ -ln($f$) which is related to 2D spin diffusion, was found above $T_c$\cite{Kambe}. In any cases, our results clearly indicate that spin correlations in LiFeP has a diffusion characteristic, meaning that spin correlation function has an anomalously large contribution at long times. Similar behavior has also been observed in La$_{0.87}$Ca$_{0.13}$FePO, but only inside the superconducting state and was suggested to be originated from a spin-triplet symmetry of superconducting state\cite{Nakai2008,Julien2008}. In our study, however, we found that the diffusive fluctuations exist far above $T_c$ in the normal state of LiFeP, indicating that they are irrelevant to superconductivity. To the best of our knowledge, the nature of spin diffusion behavior in cuprate superconductors is still unclear, although this behavior has been discovered more than two decades. Thus we hope that our work will draw more theoretical attention for this issue.

\begin{figure}
\includegraphics[width= 8 cm]{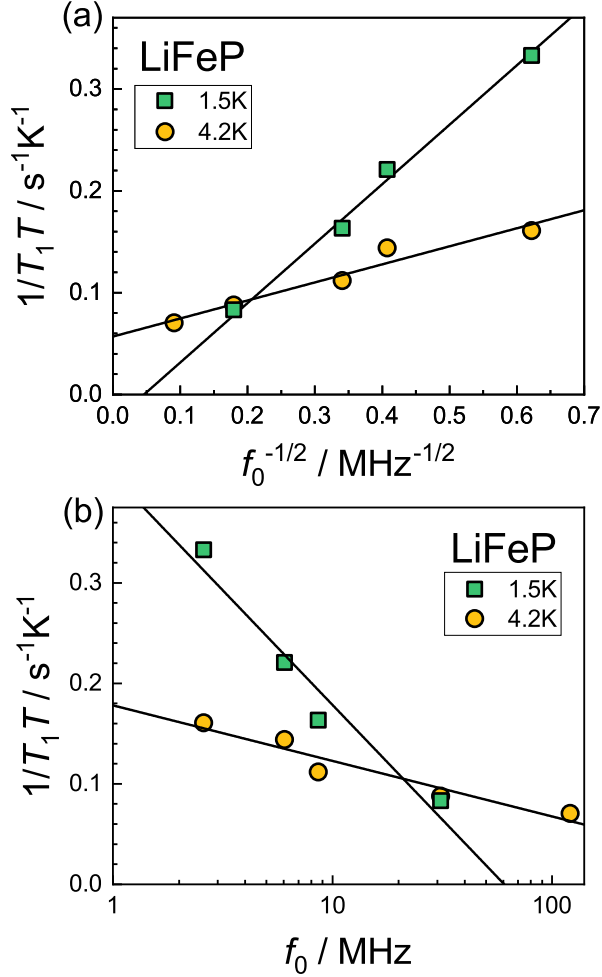}
\caption{(color online) (a) 1/$T_1$$T$ as a function of $f_0^{-1/2}$. The solid curves are the linear fittings of 1/$T_1$$T$ to $f_0^{-1/2}$. (b) 1/$T_1$$T$ as a function of the NMR frequency $f_0$. The solid curves indicate 1/$T_1$$T$ $\propto$ -ln($f$). The error bar is smaller than the symbol size.
}
\label{T1Tf}
\end{figure}

\section{Conclusion}

In summary, we investigate the superconducting gap symmetry of LiFeP and LiFeAs by London penetration depth $\lambda _{L}$ measurements. In LiFeAs, $\lambda _{L}$ is found to saturate below $T \sim$ 0.2 $T_c $, meaning that the superconducting gap is fully opened. The temperature dependence of $\lambda _{L}$ is analyzed by a two-gaps model and the value of two superconducting gaps of LiFeAs are acquired as $\Delta_1$ = 1.2 $k_B$$T_c$ and $\Delta_2$ = 2.8 $k_B$$T_c$. In contrast, we find that $\lambda _{L}$ does not show any saturation with decreasing temperature down to $T \sim$ 0.03 $T_c $ in LiFeP. This indicates the existence of nodes in the superconducting energy gap function in LiFeP. Finally, we perform spin-lattice relaxation measurements at various fields in LiFeP. 1/$T_1$$T$ is nearly temperature independent at 7 T, but is strongly enhanced at low fields below $T$ = 10 K, suggesting that the spin correlation is enhanced at very low energies. We further find that 1/$T_1$$T$ is proportional to $f^{-1/2}$, indicating that spin fluctuations have a 1D diffusive characteristic. Such behavior was also observed in some cuprate high-$T_c$ superconductors, while its origin still needs more studies.

\begin{acknowledgments}
We thank S. Kawasaki and K. Matano for assistance in some of the measurements and helpful discussions. This work was supported by NSFC (Grant No. 11904023, No. 11974405, No. 11674377 and No. 11634015) and the Fundamental Research Funds for the Central Universities (Grant No. 2018NTST22), MOST grants (No. 2016YFA0300502 and No. 2017YFA0302904), and the Strategic Priority Research Program of the Chinese Academy of Sciences (Grant No. XDB33010100).
\end{acknowledgments}

% Create the reference section using BibTeX:
%\bibliography{basename of .bib file}

\end{document}